# ARAPrototyper: Enabling Rapid Prototyping and Evaluation for Accelerator-Rich Architectures

Yu-Ting Chen, Jason Cong, Zhenman Fang, Bingjun Xiao, Peipei Zhou
Center for Domain-Specific Computing, University of California, Los Angeles
{ytchen, cong, zhenman, memoryzpp, xiao}@cs.ucla.edu

***Abstract*—Compared to conventional general-purpose processors, accelerator-rich architectures (ARAs) can provide orders-of-magnitude performance and energy gains and are emerging as one of the most promising solutions in the age of dark silicon. However, many design issues related to the complex interaction between general-purpose cores, accelerators, customized on-chip interconnects, and memory systems remain unclear and difficult to evaluate.**

**In this paper we design and implement the ARAPrototyper to enable rapid design space explorations for ARAs in real silicons and reduce the tedious prototyping efforts far down to manageable efforts. First, ARAPrototyper provides a reusable baseline prototype with a highly customizable memory system, including interconnect between accelerators and buffers, interconnect between buffers and last-level cache (LLC) or DRAM, coherency choice at LLC or DRAM, and address translation support. Second, ARAPrototyper provides a clean interface to quickly integrate users' own accelerators written in high-level synthesis (HLS) code. The whole design flow is highly automated to generate a prototype of ARA on an FPGA system-on-chip (SoC). Third, to quickly develop applications that run seamlessly on the ARA prototype, ARAPrototyper provides a system software stack, abstracts the accelerators as software libraries, and provides APIs for software developers. Our experimental results demonstrate that ARAPrototyper enables a wide range of design space explorations for ARAs at manageable prototyping efforts, which has 4,000X to 10,000X faster evaluation time than full-system simulations. We believe that ARAPrototyper can be an attractive alternative for ARA design and evaluation.**

## I. Introduction

THE scaling of conventional multicore processors has been limited by the power and utilization walls because most portions of future chips cannot be simultaneously powered up. This unpowered portion is referred to as dark silicon [1][2]. Customized acceleration [3][4][5][6][7][8][9][10][11][2] has proved to be one of the most promising solutions to address this issue. Compared to conventional general-purpose processors, these customized accelerators can provide orders-of-magnitude performance improvement and energy savings. Recently, more accelerators are being integrated into the general-purpose processors; this new architecture is referred to as the accelerator-rich architecture (ARA) [12][13][14]. Due to the significant performance and energy gains, numerous ARA efforts have been reported from both academia (such as research in [12][13][14]) and industry (such as the IBM wire-speed processors in server markets [15] and the Intel video streaming processors in consumer markets [16]).

However, the accelerator-rich architectures (ARAs) are still in the early stages of development and many design issues, especially system-level issues, remain unclear and difficult to evaluate. Examples include efficient accelerator resource management, design choices of interconnect between accelerators and scratchpad buffers, interconnect between scratchpad buffers and LLC or DRAM, efficient address translation support, etc. Therefore, a research platform that can enable rapid ARA design space explorations will be extremely useful.

In prior work, there are two major approaches used to explore the ARA design spaces: 1) full-system simulation [17][12][13][18][6][8], and 2) FPGA prototyping [19][20][5][21][9]. As shown in Figure 1, full-system simulators are very flexible when changing configurations and require little development effort to conduct design space explorations. However, the simulation time is very long and usually three to four orders-of-magnitude slower than native execution. On the other hand, FPGA prototyping provides rapid evaluation from real silicons, and it has gained increased attention. An FPGA prototype is a realization of the targeted ASIC design, which allows users to run real-life applications on the prototype at native speed and helps developers to verify the robustness of the design before taping out a chip. However, the tedious efforts for existing FPGA prototyping flows have impeded the wide adoption of FPGA prototyping for architectural design space exploration. The goal of the ARAPrototyper is to reduce the prototyping efforts to the extent that is manageable and enable both rapid prototyping and rapid evaluation/verification for ARAs.

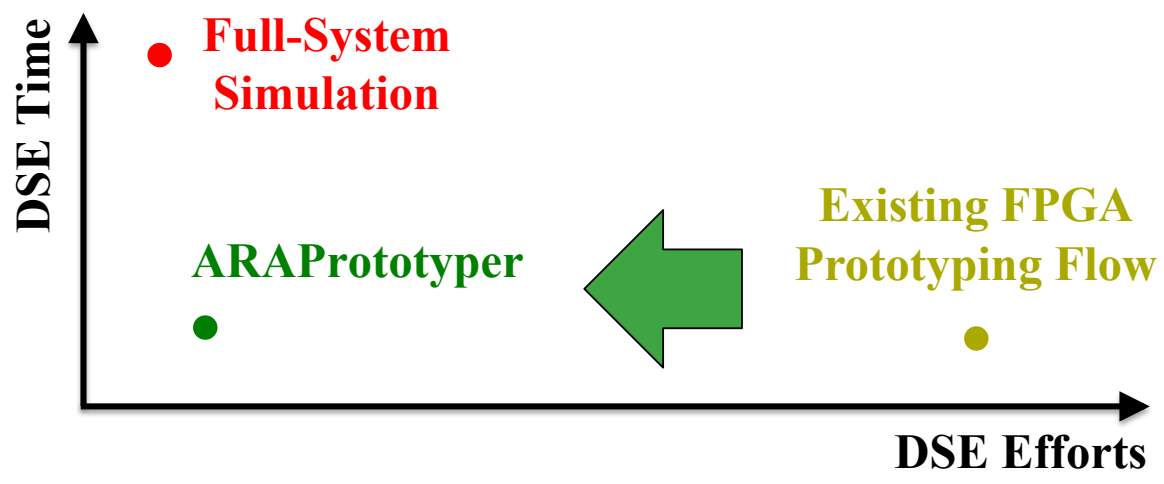

*Fig. 1:* Position of ARAPrototyper: rapid prototyping and evaluation for ARA design space exploration (DSE).

The major burden of FPGA prototyping for full-system evaluation involves significant design, implementation, and verification efforts. A robust FPGA prototype developed from scratch usually needs a very long development cycle because it requires a wide range of background knowledge, such as hardware accelerator design, system software stack support (including drivers), and application programming interfaces (APIs) design. Existing FPGA prototypes, like LegUp [22][23][21][24] and CoRAM [25][26], take years of engineering efforts for initial development and continuous improvement. An FPGA prototype developed for architectural design space exploration purposes imposes more challenges. First, architects usually want to explore different designs of ARAs or improve their ARAs in an incremental way. To reduce their burden, we should design our ARAPrototyper such that our baseline ARA prototype is highly reusable and customizable to avoid rebuilding the system from scratch. Second, users may want to add their own accelerators into the reusable baseline prototype for system-level evaluation, which still requires hundreds of lines of HLS code simply for integration in state-of-the-art FPGA prototyping flows—such as our prior effort PARC [20]. Therefore, a decent automation flow with a clean customization interface should be provided so that users can change a few lines of code and push a button to generate their own ARAs.

In this paper we present our latest prototyping flow ARAPrototyper, which enables rapid design space explorations for ARAs in native execution time. Compared to our first generation prototyping flow PARC [20], ARAPrototyper has been significantly improved (discussed in Section II-A) and has been published in two posters [27][28]. The built-in optimizer for customized interconnect between accelerators, buffers, and LLC or DRAM, has been published in [29]. In this work we choose the modern Xilinx®Zynq SoC [30], which is composed of a dual-core ARM Cortex-A9 CPU and FPGA fabrics,[1] as our underlying prototyping platform. To reduce the prototyping efforts for ARA design space explorations, we provide the following features in ARAPrototyper.

1. We develop a reusable and highly customizable baseline prototype for users to evaluate the performance of their ARAs. First, a shared memory architecture has been provided as highly parameterized hardware templates in the baseline prototype. Users can easily configure the interconnect topology between the accelerators and buffers, the interconnect topology between buffers and LLC or DRAM, coherency choice at LLC or DRAM, and TLB (translation-aside buffer) sizes in the ARA specification file without writing RTL or HLS codes. Second, to gain more insight into the performance evaluation, we add a few performance counters at the accelerator side to monitor DRAM and TLB accesses. We also leverage the existing performance counters on the ARM CPU. These can significantly reduce system design efforts and improve the quality of evaluation.
2. To further reduce the efforts of the accelerator design, we support the integration of accelerators that are written in high-level synthesis (HLS) into our ARAPrototyper. More importantly, we provide a clean accelerator integration interface for users to integrate their own accelerators by abstracting away common functionalities such as issuing memory access requests and invoking address translations. Users just have to specify a few parameters and invoke the computation kernels of their own accelerators. The whole flow to integrate users' own accelerators with their customized ARA prototype is highly automated.
3. We provide a system software stack that supports users in compiling and running their applications seamlessly on their customized ARA prototype. For users to quickly develop their applications that use those accelerators, we abstract accelerators as software libraries and provide user-friendly C/C++ APIs to manipulate accelerators.

[1]According to the Xilinx UltraScale MPSoC roadmap [31], the next generation of Zynq boards will include a quad-core ARM CPU and ultra-scale FPGA fabrics which will enable the design space explorations for even larger-scale ARAs.

To demonstrate the above design space exploration capability of the ARAPrototyper, we choose the medical imaging pipeline [32] as our main application domain for case studies. In order to illustrate the manageable prototyping efforts, we further integrate existing HLS-synthesisable accelerators from the widely used accelerator benchmark suite MachSuite [33] into ARAPrototyper. Only a few lines of code (LOCs) are needed for the integration compared to hundreds of LOCs in recent ARA prototyping work such as PARC [20]. Finally, we also compare the evaluation time of ARAPrototyper to that of the state-of-the-art full-system ARA simulator PARADE [17] by running a set of common medical imaging applications with different input sizes. ARAPrototyper achieves a 4,000X to 10,000X faster evaluation time. We believe that ARAPrototyper can be an attractive alternative to current full-system simulators for rapid ARA design and evaluation. In summary, this paper makes the following contributions:

1. Rapid FPGA prototyping for ARA design space explorations by providing a highly customizable baseline prototype with performance counters, a clean interface and automation flow to integrate the users' own HLS-synthesisable accelerators, a system software stack and accelerator APIs to quickly develop applications that can run seamlessly on the prototype.
2. Rapid evaluation of ARA designs in native execution time, which is about 4,000X to 10,000X faster than the state-of-the-art full-system ARA simulator PARADE.
3. Case studies demonstrating ARAPrototyper's capability for a wide range of ARA design space explorations, manageable prototyping efforts, and rapid evaluation time.

## II. Background and Motivation

Table I summarizes the evaluation methodologies in existing accelerator-related research. Basically, we can divide them into two major categories: simulation-based evaluation and FPGA prototyping-based evaluation.

*TABLE I:* Evaluation methodologies in existing accelerator-related research.

| Full-system | Methodology | Related work |
|---|---|---|
| N | pre-RTL simulation | Aladdin [34] |
| N | RTL simulation with SPICE models | AccStore [14] Sonic Millip3De [11] |
| N | Cycle-accurate simulation | H.264 [7] Convolution Engines [10] Conservation Cores [2] |
| Y | Full-system cycle-accurate simulation | Walker [8] DySER [6] ARC [13] CHARM [12] BiN [18] PARADE [17] |
| N | FPGA prototyping | LegUp [22][23][24] FPCA [35] CoRAM [25][26] |
| Y | Full-system FPGA Prototyping | DySER [19] TSSP [9] LINQits [5] PARC [20] |

The simulation methodologies can be further divided into the following four categories: 1) pre-RTL simulation

[34], 2) RTL simulation [14][11], 3) cycle-accurate simulation [7][10][2], and 4) full-system cycle-accurate simulation [17][12][13][18][6][8]. First, except for the pre-RTL simulation, all other simulations take a very long evaluation time that is orders-of-magnitude slower than the native execution. Second, the pre-RTL simulator Aladdin [34] uses dynamic data dependence graphs to model an accelerator, where the model depends on the input changes. More importantly, Aladdin only simulates the accelerator itself and lacks integration with full-system simulators to enable system-level exploration. Third, except for PARADE [17], all (full-system) cycle-accurate simulators also need to implement the accelerator design in RTL, which results in tedious efforts. Finally, PARADE is the state-of-the-art full-system cycle-accurate ARA simulator that provides various design space exploration choices. PARADE extends the widely used gem5 [36] simulator with HLS support to reduce the efforts of modeling the accelerators. We will compare the evaluation time of our ARAPrototyper to PARADE in Section VI-C.

Compared to the long-running simulation, FPGA prototyping [19][22][23][20][5][35][21][24][9] is gradually gaining increased attention because it enables native measurement of the performance and power in real silicons. However, the tedious prototyping efforts impede the wide adoption of FPGA prototyping for ARA design and evaluation. In this paper we exploit full-system FPGA prototyping to enable rapid design space explorations for the emerging ARAs. Our goal of ARAPrototyper is to reduce the tedious prototyping efforts far down to manageable efforts.

### *A. Comparison to Recent Prototyping Work*

In this subsection we compare the ARAPrototyper to its early version PARC [20], commercial accelerator design tools such as Xilinx SDSoC [37], and two most related prototyping work CoRAM [25][26] and LegUp [22][23][21][24].

1. **PARC [20].** PARC is our first-generation FPGA prototype designed to evaluate the ARA architecture described in [13]. ARAPrototyper shares some similar methodologies to PARC: the integration with high-level synthesis flow, shared memory architecture, and accelerator API support. However, ARAPrototyper provides many more new features. First, ARAPrototyper significantly reduces the prototyping efforts (hundreds of LOCs to a few LOCs as compared in Section VI-D) by providing a clean accelerator integration interface and automation flow. Second, ARAPrototyper significantly enlarges the scope of design space explorations for ARAs: 1) it adds the customizable interconnect layer between buffers and DRAM ports to explore the efficiency of off-chip accesses; 2) it adds the coherency choice at either LLC or DRAM. Third, ARAPrototyper adds performance counter support to provide more insights into the performance evaluation. Finally, ARAPrototyper is implemented in the newer Xilinx®Zynq SoC board [30] and has stronger ARM processor support (PARC uses a much weaker MicroBlaze processor), and thus models a real-life ARA more closely.
2. **Commercial tools [37].** FPGA vendors also provide tools to design and prototype customized SoCs. For example, designers can use Xilinx SDSoC [37] to build their own accelerators using FPGA fabrics that work together with hard ARM cores. However, it does not support most features that an ARA needs, such as the global accelerator manager, customized interconnect between accelerators, buffers, and DRAM, performance counters, to name just a few. ARAPrototyper provides a reusable baseline with highly customizable parameters for a typical ARA, and provides easy accelerator integration for rapid prototyping.
3. **CoRAM [25][26].** The goal of CoRAM is to provide a scalable and portable memory architecture so that designers can focus on the accelerator design instead of building the memory architecture from scratch. A 2D-mesh interconnect is used to provide the connectivity between CoRAM blocks, which is different from the partial crossbar architecture explored in ARAPrototyper. CoRAM provides the flexibility for designers to customize the on-chip SRAM blocks into caches, FIFOs or buffers. But designers still need to expend considerable effort to design these customized memories, which impedes the goal of rapid prototyping and evaluation. In addition, the full-system evaluation capability is not supported. Instead, ARAPrototyper provides the capability to observe interactions between hardware and OS, such as the performance impacts on TLB misses, which enlarges the scope of design space explorations.
4. **LegUp [22][23][21][24].** LegUp takes a standard C program as input and automatically compiles the program into a hybrid architecture with a MIPS soft processor and customized accelerators. The more recent update [21] uses an ARM processor in the Altera FPGA-SoC and can take OpenMP and pthread functions as input. LegUp can perform self-profiling on the processor and identify program sections that would benefit from hardware acceleration. The identified sections are synthesized by its own HLS engine. Compared to LegUp, ARAPrototyper takes a different design philosophy. ARAPrototyper allows users to design the accelerators themselves (also adopted in [23]) or leverage existing accelerators that other hardware developers provided. More importantly, ARAPrototyper models the emerging ARA architectures that have the global accelerator management (GAM), customizable interconnect between accelerators and buffers, interconnect between buffers and DRAMs, coherency choice at LLC or DRAM, etc. In addition, ARAPrototyper adds the performance counter support to provide more insights into ARA design space explorations. This is totally from a different perspective and none of these ARAPrototyper features are supported by LegUp.

## III. The Baseline ARA Prototype

We first present an overview of the ARA that we are prototyping, based on the architecture proposed in [13][18]. As shown in Figure 2, the ARA mainly contains two planes: 1) the accelerator plane, and 2) the processor plane. The accelerator plane is composed of the heterogeneous accelerators, the ARA memory system to support the high memory demand of accelerators, and IOMMU for address translation. The processor plane is composed of a conventional multicore processor with a multilevel cache. From a system perspective, the user applications are launched in the processor plane, and the compute-intensive tasks can be offloaded to the accelerator plane. The system software stack acts as the interface between the two planes. It provides the services of reservations, starts, and releases for the accelerators. The system software stack is implemented in the privileged mode and transparent to users.

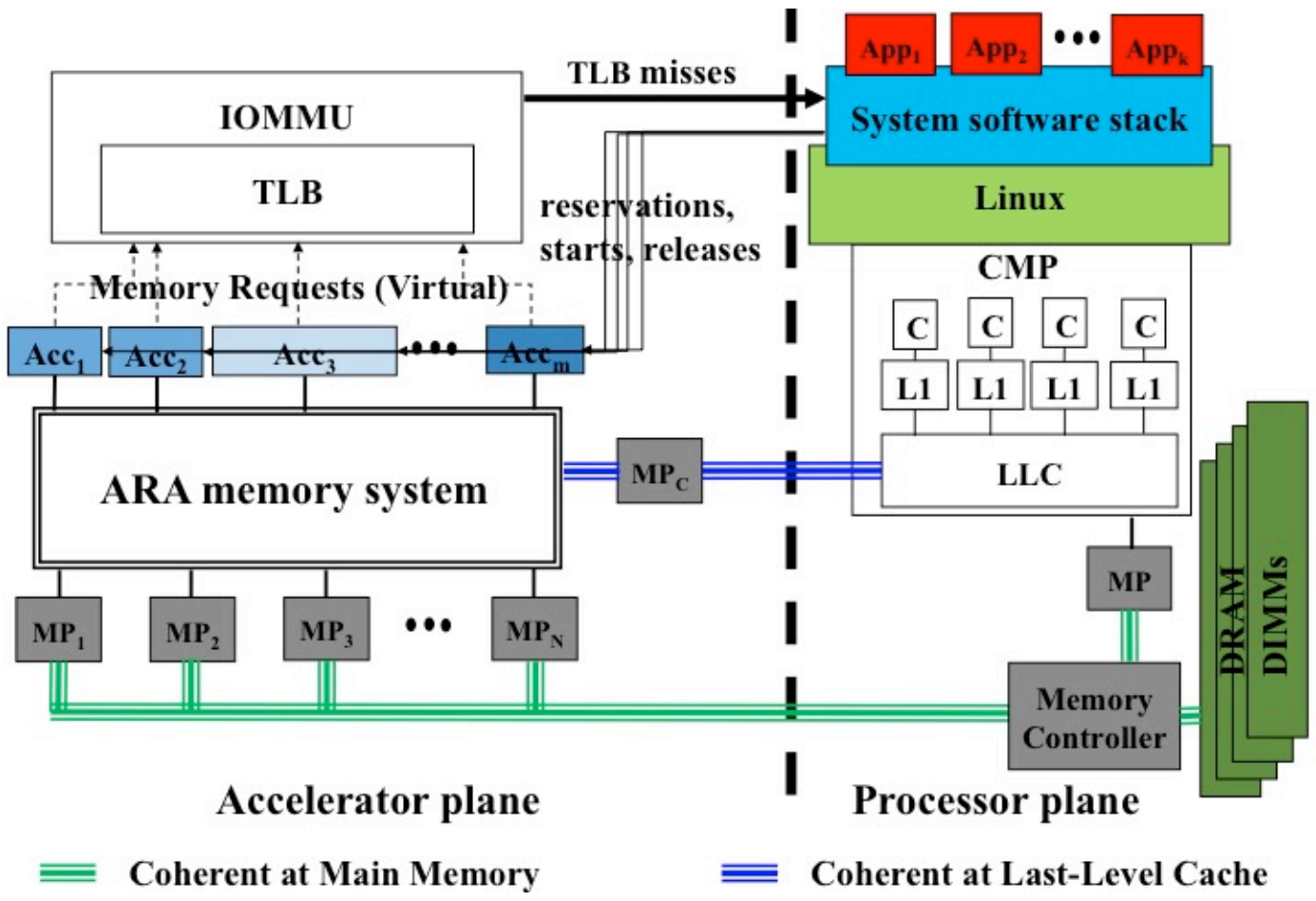

*Fig. 2:* ARA overview: accelerator plane and processor plane.

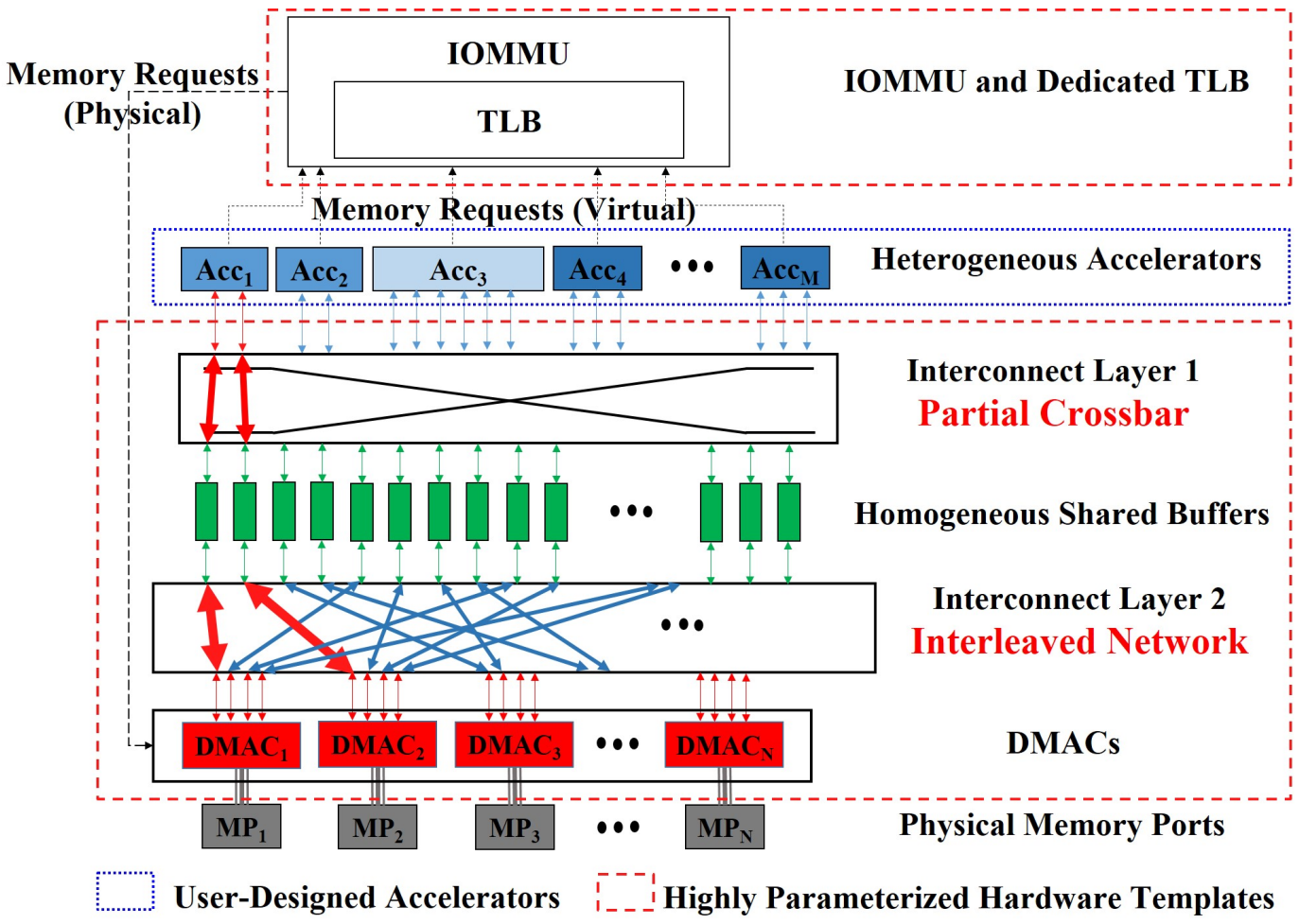

*Fig. 3:* Accelerator plane and the ARA memory system.

Next, in our baseline ARA prototype, we will present the detailed design of the customizable ARA memory system in the accelerator plane and the system software stack connecting the two planes. To gain more insights into the performance evaluation, we also add a few more performance counters in the accelerator plane, while we can leverage the existing performance counters for the processor plane. Finally, we will introduce some important features of Xilinx Zynq SoC [30], which is used for our ARA prototyping.

### *A. ARA Memory System*

The ARAPrototyper can generate a shared memory (buffer) architecture for heterogeneous accelerators to share the on-chip memory resources, which is similar to the architectures discussed in [18][14]. To share on-chip memory resources, we provide a customized two-layer interconnect, which can be synthesized automatically by specifying the parameters in the hardware templates. We also provide the flexibility if users desire to fully customize the interconnects, even to support private buffers.

Figure 3 presents the accelerator plane and its ARA memory system design in detail. The major components include 1) heterogeneous accelerators, 2) homogeneous shared buffers, 3) direct memory access controllers (DMACs), 4) physical memory ports, 5) a customized partial crossbar between accelerators and buffers, 6) a customized interleaved network between buffers and DMACs, and 7) an input/output memory management unit (IOMMU) and a dedicated TLB. We can have different types of accelerators and different numbers of accelerators for each type. Each type of accelerator has its own input and output port demands. Each port can connect to one or multiple buffers based on the generated partial crossbar topology.

The ARAPrototyper provides a pool of homogeneous buffers to be shared by the heterogeneous accelerators. Before computation, an accelerator needs to send requests to IOMMU to perform page translations. After that, IOMMU assigns corresponding DMACs to issue memory requests to fetch data through physical memory ports (MPs). The off-chip long burst requests are interleaved with the interleaved network to minimize possible conflicts. The memory requests are at the page granularity (4KB). The buffer size is 16KB by default, but can be configured by users.

*1) Customizable Optimal Partial Crossbar:* The goal of the partial crossbar is to provide sufficient connectivity between the accelerators and shared buffers. The partial crossbar avoids extra arbitration cycles that occur in a conventional bus. Therefore, a deep-pipelined accelerator can achieve initiation interval (II) as low as one with the partial crossbar support. Figure 4 demonstrates a real interconnect topology generated from ARAPrototyper. In this example, the accelerator plane contains six heterogeneous accelerators. The numbers inside each parentheses represent the assigned buffer IDs to the accelerator, which forms the topology of the customized partial crossbar. When an accelerator is reserved by an application, the accelerator has the privilege of using the assigned buffers as its own local buffers. The accelerator can fetch one element from each buffer per cycle (II = 1) since a dedicated connection is built.

The ARAPrototyper provides a built-in optimization flow [29] for the customized partial crossbar. This optimizer takes the number of ports of each accelerator and the number of shared buffers as input. Designers need to provide the maximum number of simultaneous active accelerators as another constraint. The number of simultaneous active accelerators can influence 1) the power budget and 2) the complexity of the partial crossbar, which reflect the two important design criteria—dynamic power and area. Our optimizer can guarantee the optimality of the crossbar with the minimum number of cross points based on the input and constraints. The key idea is to first guarantee the crossbar switches for those accelerators (say the number is c) with the largest memory bank demand, where one memory port exactly maps to one memory bank. For the remaining accelerators, each memory port maps to c memory banks, thus guaranteeing fully connectivity. The details of this optimizer is presented in [29] and omitted in this paper due to space constraints. In our first generation prototyping flow PARC [38], we only generate an optimal crossbar topology when the number of accelerator ports of each accelerator are equal, i.e., homogeneous. In ARAPrototyper, we have generalized the optimal partial crossbar design for accelerators with heterogeneous port demands. Finally, buffer demand information can also be reported by our built-in optimizer.

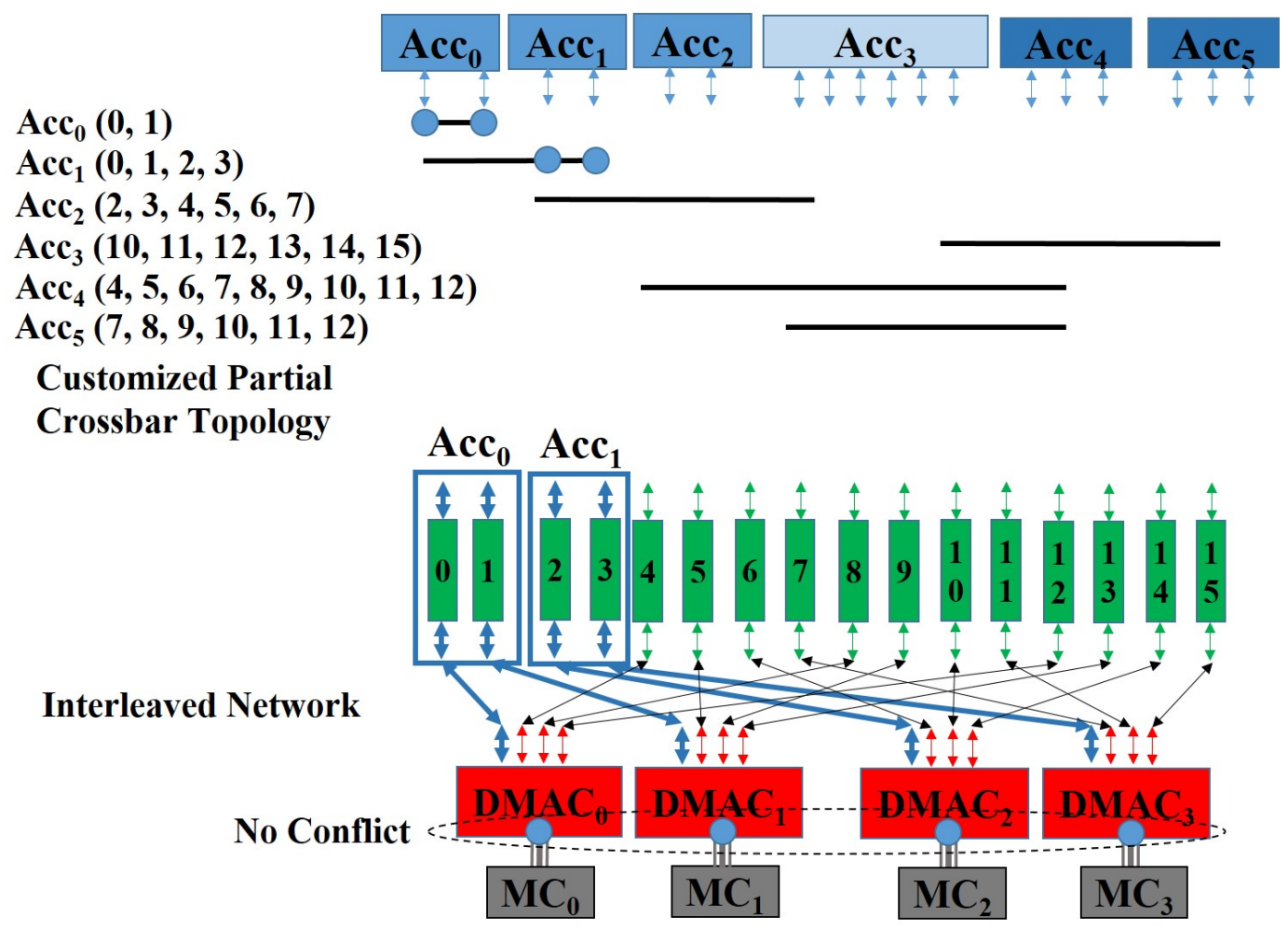

*Fig. 4:* A real example of the interconnect topology generated from ARAPrototyper.

**Private buffer architecture support**. Although we mainly target the shared buffer architecture as previously explained, ARAPrototyper also supports a private buffer architecture: each accelerator has its own buffers without sharing. Users can simply set the number of shared buffers to be equal to the number of ports of all accelerators. In this case, the shared buffer architecture is customized to the private buffers while still benefiting from the interleaved network. This can be used to evaluate an ARA with abundant buffer resources and a large power budget.

*2) Customizable Interleaved Network:* The main purpose for an interleaved network is to minimize possible conflicts for the long off-chip burst requests. An accelerator usually issues multiple requests simultaneously to prefetch data from the off-chip DRAM to on-chip buffers for near-future computation. For example, in a stencil computation, multiple data elements, e.g., five or seven, are required for a single computation. These data are prefetched into buffers in advance. If the simultaneous requests are not distributed evenly across physical memory ports, significant performance degradation can occur. First, an accelerator can start to work only when all required data are prefetched into its buffers. Second, an memory request is at the page granularity (4KB), and thus the latency is very large. The uneven distribution of requests can cause serious delay for the pending requests.

Figure 4 shows how the interleaved network successfully distributed four simultaneous accesses into four DMACs. Note that the topology of the interleaved network depends on the topology of the customized partial crossbar. In ARAPrototyper, we support two design strategies for design space exploration: 1) interleaving the requests within an accelerator, and 2) interleaving the inter-accelerator requests.

*3) Coherency Choice at LLC or DRAM:* The ARAPrototyper supports two types of coherency. Users can select either one in our flow. First, the ARA memory system can be coherent with the last-level cache (LLC) residing in the processor plane. In this case, users do not need to worry about the coherency. Second, the ARA memory system can directly exchange data with the off-chip DRAM (i.e., coherent at DRAM). Compared to the LLC coherent case, the burst DMA transfer may provide higher memory bandwidth because of larger burst sizes and more physical memory ports. However, users need to invalidate the corresponding cache lines if the data is updated in the DRAM.

*4) TLB Support in IOMMU:* Since accelerators in an ARA share the physical memory with the processor and use virtual memory for better programmability, a hardware IOMMU and a dedicated TLB is provided in the accelerator plane to support the virtual to physical address translation. The TLB size is configurable by users. We leverage the system software stack to handle a TLB miss, which will be explained in Section III-B4. To gain more insights, we also add two performance counters to monitor the TLB accesses and TLB misses. Since in our cases, the data is accessed consecutively in a streaming fashion, we can also use the TLB access counter to calculate the total DRAM accesses and achieved memory bandwidth from the accelerator plane. One can add a DRAM access counter if necessary.

### B. System Software Stack

Figure 5 presents an overview of the ARA system software stack. ARAPrototyper can automatically generate the related software modules based on the ARA specification file. The five major components in the system software stack are: 1) global accelerator manager (GAM), 2) dynamic buffer allocator (DBA), 3) coherence manager, 4) TLB miss handler, and 5) performance monitor (PM). Next we present more details of these components. Users may further customize the software stack based on their needs.

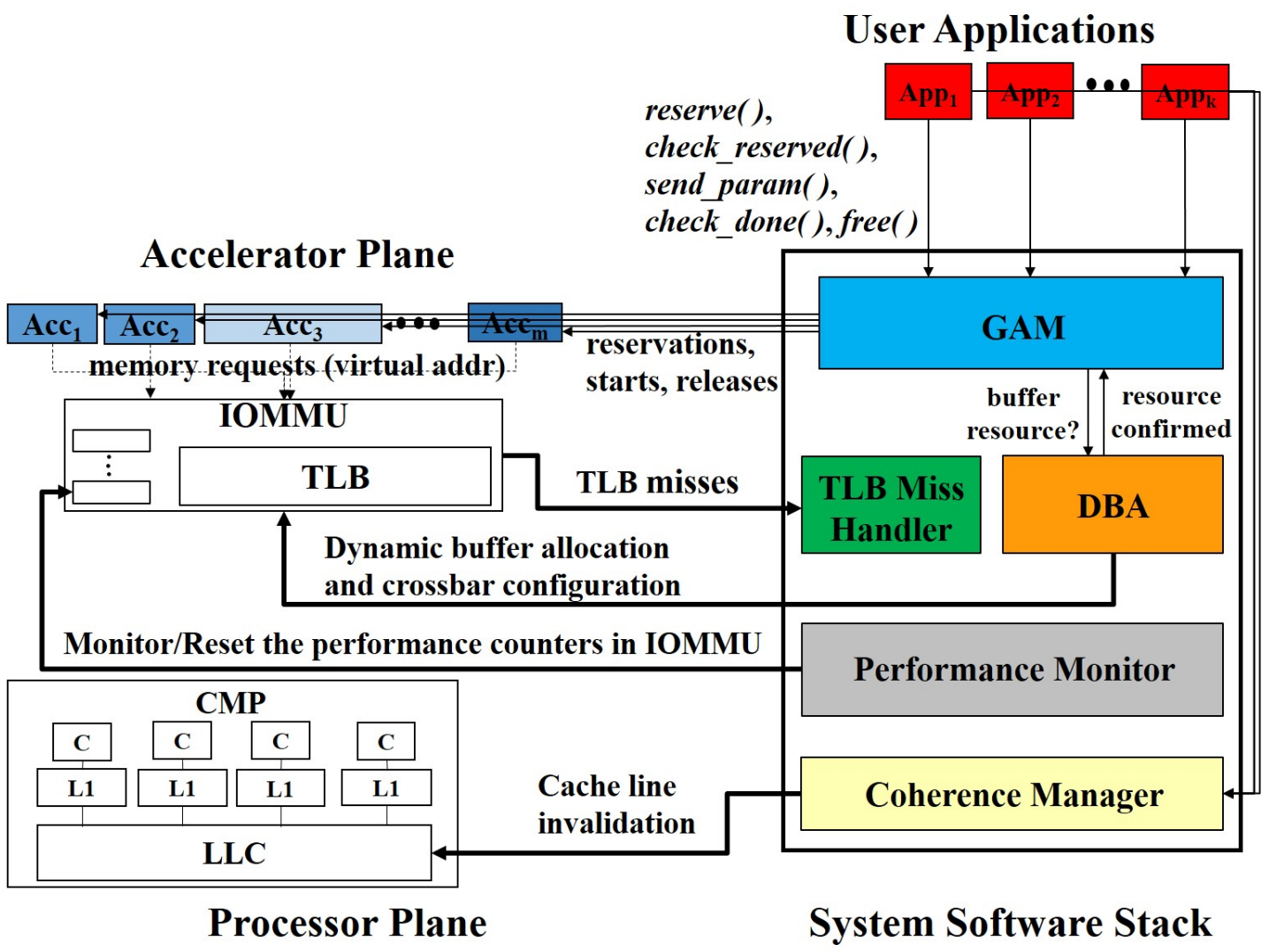

*Fig. 5:* System software stack and the interactions with the ARA and user applications.

*1) Global Accelerator Manager:* GAM is responsible for a) interfacing with user applications, b) accelerator resource management and task scheduling, and c) requesting buffer resources. User applications can talk to GAM with the provided APIs, which will be discussed in Section V. In GAM, we use a table to keep track of the available accelerators of each type. The incoming requests from user applications are scheduled in a first-come, first-serve basis. GAM would make requests for the shared buffer resources to the dynamic buffer allocator before reserving a target accelerator. As observed in our PARC [20] work, GAM runs on top of a separate ARM CPU core and is sufficient for managing FPGA accelerators used in our prototype. This design is easier than the dedicated hardware GAM for managing ASIC accelerators used in PARADE [17].

*2) Dynamic Buffer Allocator:* In the shared memory/buffer architecture, discussed in Section III-A, a buffer bank can be shared by multiple accelerator ports. DBA is in charge of the dynamic buffer assignment during runtime based on the requests from user applications. Static assignment, which is used in our first-generation prototyping flow PARC [20], can no longer handle the dynamic cases and can limit the framework scalability for evaluation.

DBA receives the buffer requests from GAM. As shown in Figure 6, DBA uses a list structure, called *task list*, to store the requests that have not been processed. With information of the incoming tasks, DBA is able to provide different kinds of allocation policies to influence task scheduling. Users are able to modify the allocation policy of DBA based on their demand, such as throughput-driven or deadline-driven scheduling, to manipulate the *task list*.

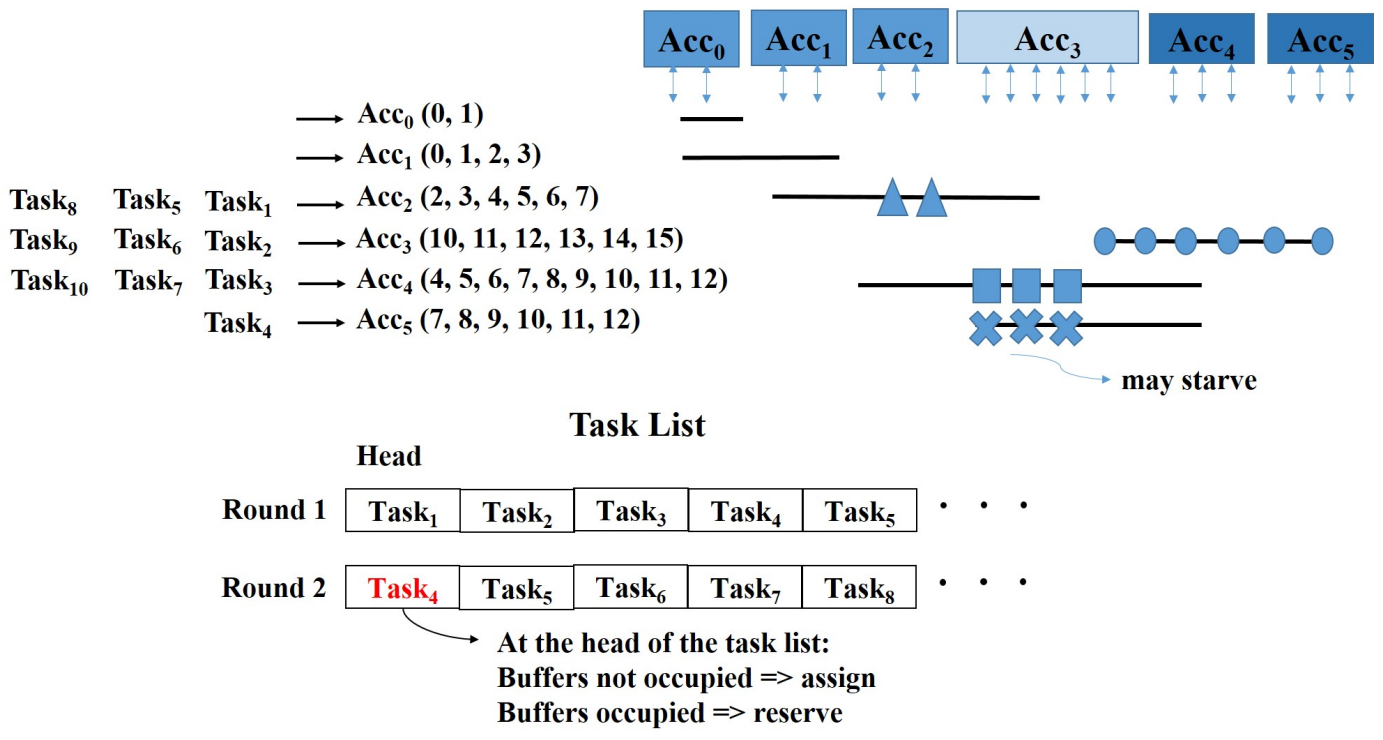

*Fig. 6:* Dynamic buffer allocation: a starvation-free scheme.

In ARAPrototyper, we provide a starvation-free buffer allocation policy, as illustrated in Figure 6. The tasks come in numerical order. It is possible that $Acc_5$ can starve since $Acc_2$, $Acc_3$, and $Acc_4$ occupy the buffers to serve the continuous incoming tasks. To prevent starvation, we use two flags for each buffer: occupied and reserved. A buffer can only be allocated by DBA when the buffer is neither occupied nor reserved. When a buffer is assigned to an accelerator, it will be marked as occupied. The reserved flag is used when a buffer is occupied but another accelerator would like to reserve it. The starvation can be resolved by providing the "reserved" privilege only for the task at the head of the *task list*. This guarantees the task at the head can always occupy or reserve the required buffers. After that, the algorithm greedily allocates buffers to the tasks in the *task list* in order until no feasible allocation can be found. This algorithm is lightweight and the overhead is negligible.

*3) Coherency Manager:* The ARAPrototyper offers a coherency manager for coarse-grained software-based coherence handling. When users try to directly write data to DRAM for higher memory bandwidth, the overlapping pages residing in multilevel caches in the processor need to be invalidated. We abstract the cache invalidation details in the coherency manager in our system software, so users only need to call the coherency manger to handle the possible coherency issue.

*4) TLB Miss Handler:* For the TLB misses arising from the accelerators, we currently use a software-based handler to handle the miss. To reduce overhead incurred in the communication between IOMMU and the TLB miss handler in the privileged mode, IOMMU groups multiple TLB misses and sends them to the handler together. Instead of using the slow kernel API to do page translation, we write our own version by leveraging the ARM architecture support for page table walk. Table II shows our profiling results on the average TLB miss handling time. Our efficient walker reduces the miss penalty from 4278 cycles to 458 cycles, and thus overhead from TLB misses in an ARA can be significantly reduced.

*TABLE II:* Average TLB miss penalty; kernel APIs vs. software page table walk (Acc@100MHz).

| | Microblaze (PARC [20]) | Cortex-A9 | Cortex-A9 |
|---|---|---|---|
| Method | Kernel APIs | Kernel APIs | pgtwalk |
| Freq. | 100MHz | 667MHz | 667MHz |
| Cycles | 4975 | 4278 | 458 |
| Time(us) | 49.75 | 6.41 | 0.69 |

We are also considering implementing a hardware-based page walker. It can be a scalable version when the number of accelerators is large. However, the hardware-based walker in our Zynq board [30] can lead to three sequential DRAM accesses (600 cycles) per TLB miss because of the walk on multilevel page tables.

*5) Performance Monitor:* To provide more insights into the system bottleneck analysis, we add performance counters in the IOMMU so that the TLB hit/miss events and memory bandwidth can be monitored on-the-fly, as mentioned in Section III-A4. We add a PM module in the system software stack to handle requests from applications and interact with the IOMMU to monitor or reset the performance counters. These performance counters can provide more in-depth performance characterization in addition to simple runtime numbers. The impact of different architecture parameters can also be observed and analyzed.

In addition to using the accelerator-side performance counters supported by PM, users can also use OProfile [39] to obtain the performance counter information inside CPU cores. We have successfully ported OProfile on top of our ARA baseline prototype under the Zynq platform.

### C. Prototyping Platform: Xilinx Zynq SoC

We choose the Xilinx Zynq ZC706 evaluation board [30] with 1GB DRAM as our underlying prototyping platform. Figure 8 shows the architecture of Zynq SoC. It is composed of FPGA fabrics for accelerator implementation and a dual-core ARM for the system software implementation. The FPGA contains around 2MB on-chip block RAM, which can be used to implement the shared buffers. User applications can be launched on the ARM processor with Linux support. The Zynq architecture has the following advantages to support ARAPrototyper.

1. **Faster processor cores.** The hard ARM cores can run up to 800MHz, which is much more efficient than the soft Microblaze cores synthesized from FPGA. Linux can be ported on ARM cores and executed fluently. The system software stack provided in ARAPrototyper, including global accelerator manager, dynamic buffer allocator, coherence manager, TLB miss handler, and performance monitor, can all leverage the faster processor.
2. **A coherent LLC support.** The dual-core ARM provides a shared L2 cache. The shared buffers can be coherent with L2 cache through the accelerator-coherent port (ACP). This provides an alternative ARA design opportunity (as described in Section III-A3).
3. **A fast ASIC on-chip memory controller.** The ASIC on-chip memory controller provides higher memory bandwidth compared to a memory controller synthesized in FPGA. In order to efficiently exploit the available memory bandwidth, Zynq provides four high-performance (HP) ports in the FPGA fabric. This gives us opportunities to explore the topology of the interleaved network to better utilize the off-chip memory bandwidth.

Our prior PARC work [20] is prototyped on the ML605 board with Virtex 6 FPGA. Compared to the Virtex 6 with only FPGA fabrics, Zynq SoC enables a wider range of ARA design explorations.

## IV. Design Automation Flow and ARA Customization Interface

The main challenge to do architectural design space exploration through FPGA prototyping is the long development cycle for each generation of an ARA, which requires extensive coding in RTL. To further reduce the prototyping efforts, we develop a highly automated design flow for users to customize the baseline ARA prototype and integrate their own accelerators. Users only have to configure an ARA specification XML file to customize their ARA, and specify a few parameters in the acceleration integration interface to add their own accelerators that are written in HLS. Our design flow can automatically generate users' customized ARAs and deploy them on the underlying FPGA prototyping platforms.

### A. Design Automation Flow

We classify the components in the ARA prototype into the following three categories, as shown in Figure 7.

1. **Platform-specific modules** are mainly bonded to the hard modules in the FPGA chip and evaluation platform (board), such as the dual-core ARM processor. ARAPrototyper can adapt to different platforms, and thus users can spend minimum effort on the platform issues.
2. **Platform-independent modules** include two-layer interconnects, shared buffers, IOMMU and TLB, and DMACs, which are the major components in the ARA memory system. ARAPrototyper provides highly parameterized hardware templates for these platform-independent components. Users can easily customize them in the ARA specification file that will be explained in Section IV-B.
3. **User-designed accelerators.** In the ARAPrototyper, we provide a group of highly optimized accelerators in the medical imaging pipeline that will be further explained in Section VI. Users can easily develop their own accelerators in HLS. Furthermore, we provide a clean accelerator integration interface (to be explained in Section IV-C) for users to easily add their own accelerator into our ARAPrototyper.

Figure 7 presents our design automation flow, which enables ARAPrototyper to maximize the reuse of hardware modules and user-designed accelerators when developing ARAs. It begins with the ARA specification file, and all following steps can be executed automatically upon a single "make" button. In the left branch, we apply ARA memory system optimizations to our platform-independent modules by using the configurations in the ARA specification file. This is combined with our hardware templates to create the ARA memory system. In the middle branch, HLS tools are applied to user-designed accelerators coded in C/C++ for generating RTL designs. In the right branch, the platform information is used to generate platform-specific modules. Depending on the target platform, e.g., Xilinx Zynq FPGA in our case, the flow can be seamlessly integrated with the corresponding back-end process (e.g., Xilinx PlanAhead flow for bitstream generation).

### B. ARA Specification File

```
<system>

<ACCs>
   <acc type="gradient" num="2" num_params="5">
       <port size="16K" num="6"/>
   </acc>
   <acc type="segmentation" num="1" num_params="13">
       <port size="16K" num="8"/>
   </acc>
   <acc type="rician" num="1" num_params="7">
       <port size="16K" num="12"/>
   </acc>
   <acc type="gaussian" num="1" num_params="7">
       <port size="16K" num="5"/>
   </acc>
</ACCs>

<SharedBuffers size="16K" num="32" numDMACs="4"/>

<Interconnects>
   <ACCs_to_Buffers type="crossbar" connectivity="3" auto="1"/>
   <Buffers_to_DMACs type="interleaved" use="1" auto="1"/>
</Interconnects>

<IOMMU>
   <TLB size="8K" evict="LRU"/>
<IOMMU/>

<CoherentCache use="0" />

<AccFrequency hz="100MHz" />

</system>
```

*Listing 1:* An example ARA specification file created for design space exploration with four types of accelerators via ARAPrototyper.

The ARA specification file is provided for users to specify components and configure the parameters in the accelerator plane. Users can easily evaluate their new accelerators by integrating them into the reusable baseline prototype, and perform system-level design space explorations. It is composed of six major sections: (1) accelerator specification, (2) shared buffer and DMAC specification, (3) interconnect specification, (4) IOMMU specification, (5) coherence specification, and (6) target frequency. The ARA specification file is recorded in a XML format and users can easily modify it on top of the XML template we provide, including all the six sections.

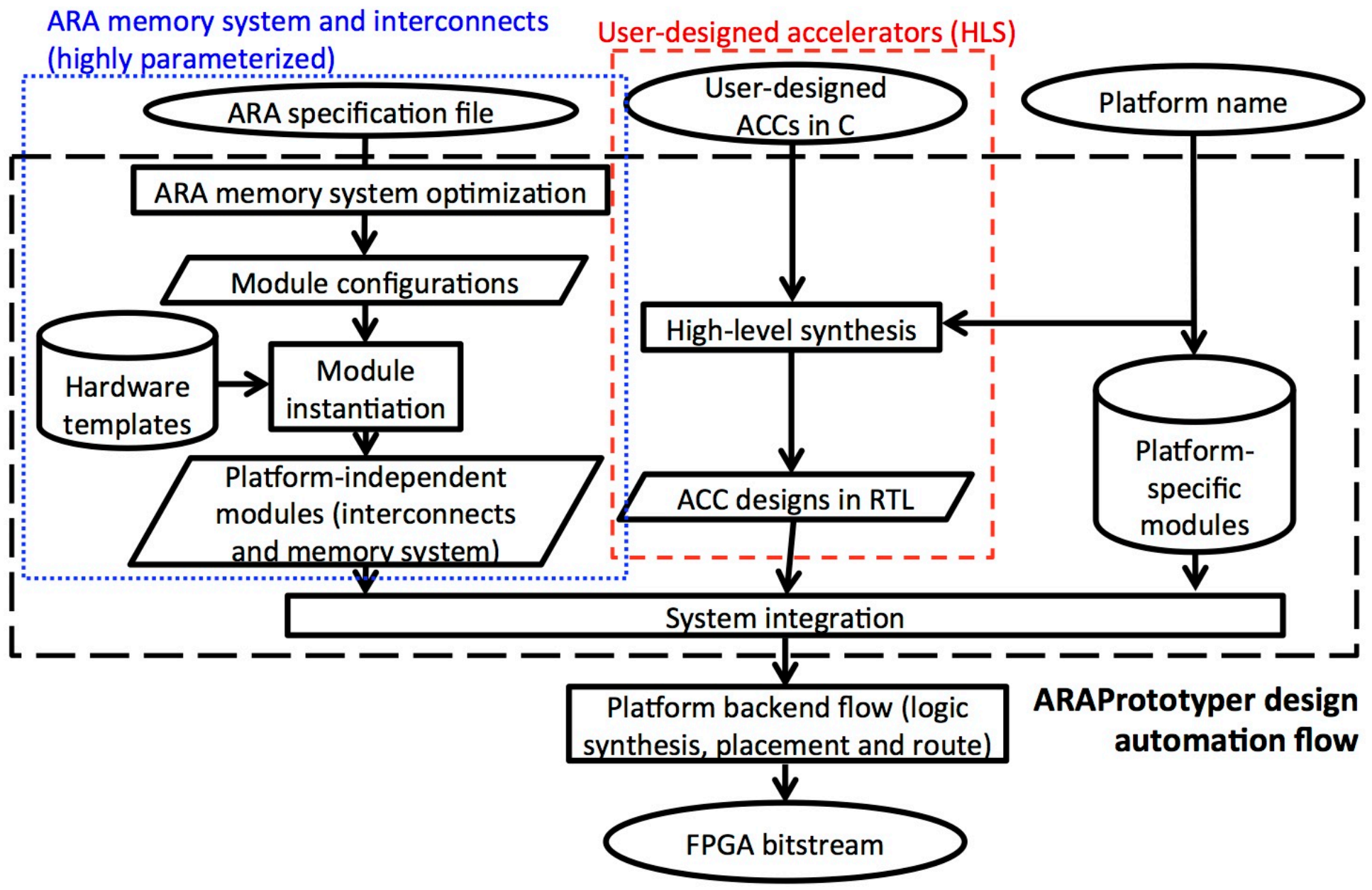

*Fig. 7:* ARAPrototyper design automation flow.

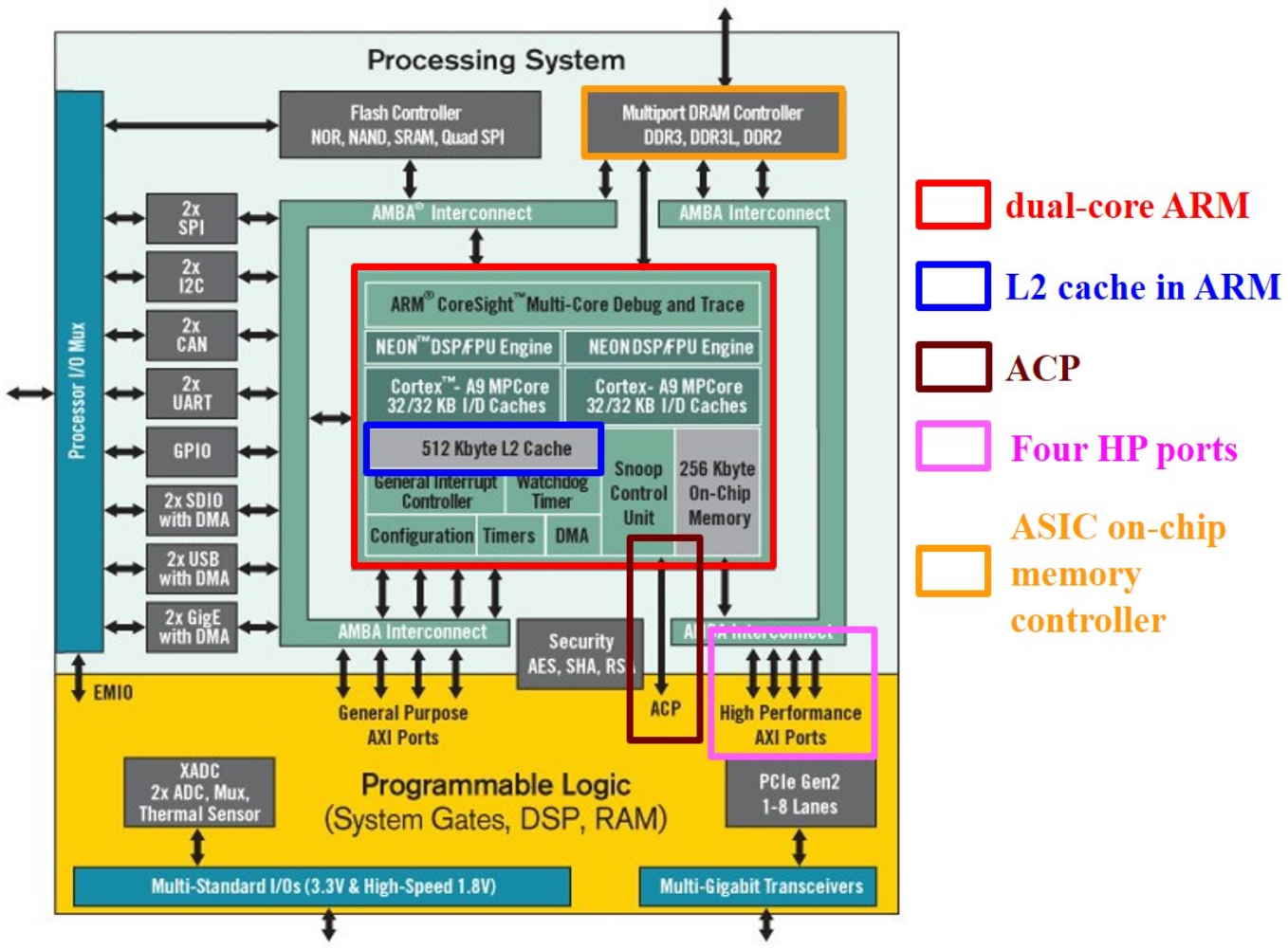

*Fig. 8:* The prototyping platform: Xilinx Zynq SoC.

Listing 1 is an example of an ARA specification file. In section "ACCs," users can specify the name, the number of duplications, the number of input parameters, the number of ports (buffers) and the needed buffer size of each type of accelerator. In section "SharedBuffers," the total number of buffers, the size of a buffer, and the number of DMACs, can be configured. In section "Interconnects," ARAPrototyper allows users to configure the interconnects between accelerators and shared buffers, and the interconnects between shared buffers and DMACs. The "connectivity=3" means the generated partial crossbar can guarantee a feasible crossbar configuration for at least any three accelerators to have their own buffer resource at a given time in the ARA. The "auto" field is set to "1" to use the built-in optimizer. ARAPrototyper also provides the flexibility for users to specify the interconnects by themselves. In section "IOMMU," the users can configure the TLB size and eviction policy. In section "CoherentCache," users can specify whether the coherent L2 cache in the dual-core ARM is used. In section "AccFrequency," the clock frequency in the accelerator plane can be configured. As long as this file is prepared, the ARAPrototyper automation flow can be invoked for automated system generation. Compared to our earlier flow PARC [20], we keep section (1) and (2), but significantly enrich the configurations of the shared memory architecture and other design features in section (3) to (6).

### C. Accelerator Integration Interface

To further reduce the development cycle of adding users' own accelerators, ARAPrototyper supports the integration of accelerators developed in HLS. Moreover, ARAPrototyper provides a clean accelerator integration interface that only needs a few lines of code to integrate users' own accelerators.

**Accelerator development in HLS.** Design productivity of accelerators can be significantly improved by raising the level of design abstraction beyond RTL. HLS tools [40][41] enable automatic synthesis of high-level, untimed or partially timed specifications (such as in C, C++, or SystemC) to low-level cycle-accurate RTL code. As reported in [40], the code density can be easily reduced by 7 to 10X when moved to high-level specification in C, C++, or SystemC; and at the same time, resource usage can also be reduced by 11% to 31% in an HLS solution, compared to a hand-coded RTL design.

**Accelerator integration interface.** To allow users to easily integrate existing accelerators into the ARAPrototyper, we provide the following flow. First, designers need to specify the accelerator port information in the ARA specification file, such as the number of parameters sent from CPU and the number of demanded buffers. Next, our tool can generate the

```
void vecsqr (volatile int* vaddr_port0, volatile int* vaddr_port1,
   volatile int* buf_id_port0, volatile int* buf_id_port1,
   ...
   volatile int* IOMMU_FIFO,
   volatile float port0[CHUNK_SIZE],
   volatile float port1[CHUNK_SIZE]
   ...) {
      // read function parameters
      int vaddr0  = *vaddr_port0;
      ...
      // issue read memory access
      memory_request0(READ, IOMMU_FIFO, vaddr0, buf_id0, Req_Length0);
      ...
      // computation kernel to be plugged in
      for(int i = 0; i < Req_Length1; i ++)
         port1[i] = port0[i] * port0[i];
      ...
      // issue write memory access
      memory_request1(WRITE, IOMMU_FIFO, vaddr1, buf_id1, Req_Length1);
}
```

*Fig. 9:* Accelerator integration template in HLS-compatible C.

port names and the corresponding HLS pragma for the control and data ports between the accelerator, IOMMU, and CPU automatically using the ARA specification file. This generated file in HLS-compatible C format is called accelerator integration template, as shown in Figure 9. Designers need to place the computation kernel and the invoking of read and write memory requests explicitly in the corresponding locations in the template. After that, our flow can automatically generate ARAPrototyper-compatible HLS codes.

Control and data ports, are generated as function parameters in the HLS codes. There are three kinds of ports. First, the input parameters sent from CPU (such as "vaddr_port0") are generated. These parameters are sent from the CPU through the AXI-Lite port and are stored in the registers of the accelerator. Second, the communication channel ("IOMMU_FIFO") realized by FIFO is generated. The accelerator uses the FIFO to send read and write requests to IOMMU to fetch data from DRAM (or L2 cache) to its own buffers and write data back from its own buffers to DRAM (or L2). Third, the ports to input and output buffers such as "port0" and "port1" are generated. These ports are connected to the shared buffers through the automatically synthesized crossbar described in Section III-A1.

The only changes that designers need to specify are the memory requests in reading data from DRAM (or L2 cache) and writing data back to DRAM (or L2) in the accelerator integration template. For example, a memory request ("memory_request0") needs to be specified explicitly before the computational kernel reads data from its own buffer. Similarly, the output results need to be written back after computation is done in "memory_request1." This only involves a few lines of code (LOCs) change. As shown in Figure 9, after the existing accelerator computation kernel is plugged in, only the two lines with "Req_Length0" and "Req_Length1" (shown in red color and bold italic font) are added. Detailed prototyping efforts (LOCs) will be presented in Section VI-D.

## V. Application Development API

To enable rapid development of applications that use the accelerators in ARAPrototyper, we abstract accelerators as software libraries and provide the user-friendly C/C++ APIs.

```
#include "accelerator_type.h"
void main() {
   class Acc_Gaussian acc; Image a;
   acc.reserve();
   while(acc.check_reserved() == 0);
   acc.send_param(7, Image::get_M(), Image::get_N(),
      Image::get_P(), a.get_ptr(), 1, 1, 1);
   while(acc.check_done() == 0) wait(1000);
   acc.free();
}
```

**(a)**

```
#include "accelerator_type.h"
void main() {
   class Acc_Gaussian acc;  Image a;
   acc.run(7, Image::get_M(), Image::get_N(),
      Image::get_P(), a.get_ptr(), 1, 1, 1);
}
```

**(b)**

```
#include "accelerator_type.h"
#include "TLB_PM_api.h"
void main() {
   class Acc_Gaussian acc;  Image a;
   class TLB_Performance_Monitor pm;
   pm.reset_tlb_counters();
   acc.run(7, Image::get_M(), Image::get_N(),
      Image::get_P(), a.get_ptr(), 1, 1, 1);
   int access_num = pm.get_tlb_access_num();
   int miss_num = pm.get_tlb_miss_num();
}
```

**(c)**

*Fig. 10:* Code examples of using APIs to develop applications.

With the information provided by the ARA specification file, ARAPrototyper can automatically generate the header file of accelerator APIs for programmers, which is similar to [20]. For each type of accelerator, we provide the following APIs with fine-grained accelerator control: 1) *reserve()*, 2) *check_reserved()*, 3) *send_param()*, 4) *check_done()*, and 5) *free()*. Users can develop their applications with the C++ classes and member functions to manipulate the accelerators in the ARA. After applications are developed, users can simply set up a g++ cross compiler to compile applications into ARM executable, which can be seamlessly executed on the ported Linux on the Zynq board.

Figure 10(a) is an application example using an accelerator. It first uses the *reserve()* function to make requests to GAM for reserving an accelerator. After the reservation is confirmed, the required parameters are sent through the *send_param()* function and the accelerator will be started. The application should periodically check the status of accelerators with *check_done()*. Once the accelerator finishes its task, it should be freed for future use. With these APIs, a programmer can explore more complicated settings, such as using multiple accelerators simultaneously in the user application.

Moreover, our latest ARAPrototyper adds a simplified API, called *run()*, which is intended for software developers who do not want to dig into the hardware accelerator details. This API covers the functionality from reserving to releasing the accelerator using a single function. Figure 10(b) is a code example, which achieves the same functionality that the code of Figure 10(a) does.

As presented in Section III-B5, ARAPrototyper provides a PM module to monitor performance counters added in our prototype. We provide several APIs built on top of the PM module so that designers can use these APIs to monitor those key performance counters for analyzing and improving the ARA design. Figure 10(c) shows how TLB accesses and misses can be monitored by using the provided APIs.

## VI. Experimental Results

In this section we present a quantitative evaluation of the rapid evaluation time and manageable prototyping efforts of ARAPrototyper for ARA design space explorations. We choose the medical imaging processing pipeline as our target application domain, where customized accelerators can achieve orders-of-magnitude energy gains. To demonstrate the capability and usage of ARAPrototyper, we conduct a number of case studies for ARA design explorations.

### A. Target Domain: Medical Imaging

The target application domain we choose to accelerate is primarily the medical imaging processing pipeline, which is one of the most important application domains in personalized medical care. This pipeline is used to process the raw data obtained from computerized tomography (CT) [32]. We are motivated to accelerate it with a scenario in which a doctor is able to show the CT analysis interactively to patients with a tablet. Current mobile processors without accelerators cannot provide real-time and energy-efficient solutions.

The medical imaging pipeline can be divided into the following stages. After image reconstruction, it would 1) remove noise and blur; 2) align the current image with previous images from an individual; and 3) segment a region of interest for diagnosis [32]. The three tasks can be implemented by four accelerator kernels: $gradient$, $gaussian$, $rician$, and $segmentation$. By default, we use 128 slices of 128*128 images as the kernel input. In the following subsections, we will mainly demonstrate the benefits of ARAPrototyper using these four accelerator kernels for case studies.

### B. Overall ARA Performance and Energy-Efficiency

First of all, we present the overall performance and energy-efficiency of ARA. Table III presents the performance, power, and energy-efficiency results of the aforementioned four medical imaging kernels on state-of-the-art Intel Xeon and ARM processors, our ARA FPGA prototype, and projected ARA on ASIC. We use OpenMP to implement a 24-thread parallel version of each kernel on the Intel Xeon processor. The result on ARM Cortex-A9 uses a single thread. All kernels are compiled using gcc with -O2 option. Note that for the ARA version, in this experiment we only put a single processing element (PE) for each kernel (except for gradient which has two PEs), which can be easily duplicated. As shown in Table III, such an ARA FPGA prototype can achieve 3.9X to 65.4X better energy-efficiency than the 24-thread Intel Xeon CPU. According to the report in [42], the power gap between FPGA and ASIC is around 14X and delay gap is around 3.4X to 4.6x. If our ARA is implemented in an 45nm ASIC, 217X-3,661X energy savings over the Intel Xeon CPU are expected.

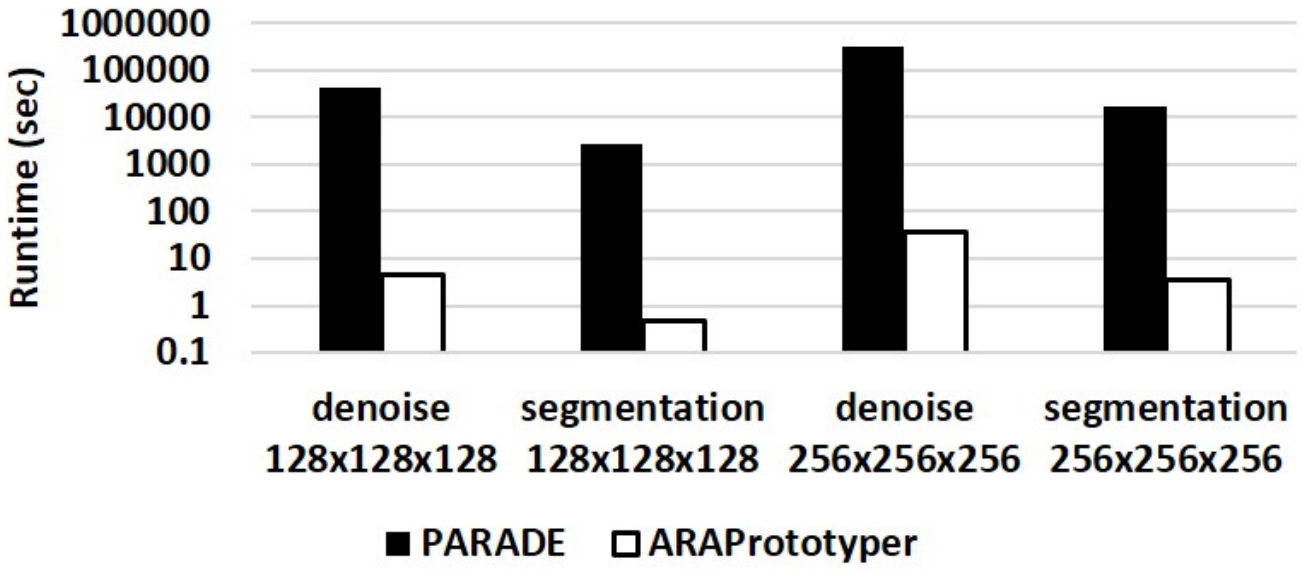

*Fig. 11:* Evaluation time on ARAPrototyper and PARADE [17].

### C. Evaluation Time

To demonstrate the rapid evaluation time of ARAPrototyper, we compare it against the state-of-the-art full-system ARA simulator PARADE [17] for a typical ARA configuration. Figure 11 compares the execution time of two common medical imaging applications with different input sizes on ARAPrototyper and PARADE. For the larger input size, it takes a couple of days for PARADE to simulate one single ARA configuration; while it only takes a minute or so for ARAPrototyper to run the configuration. We should mention that our flow generation time (generating the ARA configuration to the FPGA bitstream) takes around four hours, but it is a one-time effort for one ARA configuration that can run multiple applications with multiple inputs. Usually, the native executions on our FPGA prototype are 4,000X to 10,000X faster than full-system simulations, and we believe ARAPrototyper can be an attractive alternative for design space explorations.

### D. Prototyping Efforts

To demonstrate the manageable prototyping efforts of ARAPrototyper, we present the lines of code (LOCs) that users have to change or add to customize their own ARA using existing accelerators in the prototype or integrating their own accelerators.

We first present LOCs for users to configure their own ARA by leveraging our reusable baseline prototype and existing accelerators in the prototype. As shown in Table V, users can simply configure the ARA specification file with up to 33 lines of XML code to set up the parameters of the shared memory architecture and operating frequency. There is no C/C++ description or RTL code required. Users can start the push-button ARAPrototyper flow after the specification file is set to obtain FPGA prototype in hours. We also present the LOCs for automatically generated RTL from the baseline prototype in Table V, which is more than 37,000 lines. It reflects the huge engineering efforts required if everything is built from scratch.

Next, we present LOCs for users to integrate their own accelerators. To demonstrate the benefits of reduced prototyping efforts of ARAPrototyper compared to our first generation prototyping flow PARC [20], we also include it for a quantitative comparison. Table IV presents the LOCs to integrate our medical imaging accelerators and third-party MachSuite [33] accelerator kernels into PARC and ARAPrototyper, including total generated RTL code, total HLS C/C++ code, kernel-only HLS code, and integration-only code. We include eight more accelerator kernels from a widely used third-party accelerator benchmark suite MachSuite to better illustrate our manageable prototyping efforts.[2] As shown in Table IV, compared to the hundreds of LOCs for accelerator integration in PARC, users only need to add a few LOCs (most of the time less than 10 LOCs) to integrate their own accelerators into ARAPrototyper due to its clean accelerator integration interface and automation flow.

*TABLE III:* Performance and energy comparison over 1) Intel Xeon (Haswell), 2) ARM Cortex-A9, 3) ARA FPGA prototype, and 4) projected ARA on ASIC

| | | Xeon | Cortex-A9 | ARA on FPGA | ARA on ASIC |
|---|---|---|---|---|---|
| Config. | Freq. | 24 threads 1.9GHz | 667MHz | Acc@100MHz CPU@667MHz | — 45nm |
| | Power | 190W(TDP) | 1.1W | 3.1W | 0.22W |
| gradient | Runtime(s) | 0.030 | 2.48 | 0.33 | 0.08 |
| | Energy Eff. | 1x | 2.07x | 5.56x | 311x |
| rician | Runtime(s) | 0.036 | 2.60 | 0.57 | 0.14 |
| | Energy Eff. | 1x | 2.39x | 3.89x | 217x |
| gaussian | Runtime(s) | 0.14 | 6.31 | 0.34 | 0.08 |
| | Energy Eff. | 1x | 3.87x | 25.59x | 1432x |
| segmentation | Runtime(s) | 0.40 | 13.67 | 0.37 | 0.09 |
| | Energy Eff. | 1x | 4.03x | 65.38x | 3661x |

*TABLE IV:* Lines of code (LOCs) to integrate medical imaging and third-party MachSuite kernels into PARC/ARAPrototyper, including total generated RTL code, total HLS C/C++ code, kernel only HLS code, and integration-only code.

| Domain | Accelerator | Total RTL | Total HLS | Kernel | PARC Integration | ARAPrototyper Integration |
|---|---|---|---|---|---|---|
| Medical Imaging | gaussian | 15107 | 513 | 363 | 150 | 5 |
| | gradient | 32538 | 778 | 616 | 162 | 6 |
| | segmentation | 63857 | 1304 | 1070 | 234 | 8 |
| | rician | 42291 | 1140 | 850 | 290 | 12 |
| MachSuite [33] (third-party) | FFT/TRANSPOSE | 17072 | 530 | 412 | 118 | 4 |
| | GEMM/NCUBED | 3201 | 121 | 23 | 98 | 3 |
| | GEMM/BLOCKED | 5226 | 158 | 20 | 138 | 5 |
| | KMP/KMP | 3593 | 167 | 45 | 122 | 4 |
| | MD/KNN | 7023 | 243 | 53 | 190 | 7 |
| | SORT/MERGE | 2996 | 128 | 54 | 74 | 2 |
| | SPMV/CRS | 4080 | 160 | 18 | 142 | 5 |
| | VITERBI/VITERBI | 4212 | 177 | 35 | 142 | 5 |

*TABLE V:* Lines of code (LOCs) to customize users' own ARA prototype using existing accelerators.

| | | # line of XML code |
|---|---|---|
| input | ARA description file | 33 |
| | components | # line of RTL code |
| automatically generated | IOMMU | 21407 |
| | crossbar | 1526 |
| | top module | 14253 |
| | total | 37186 |

[2]We do not include MachSuite accelerators for more studies because their performance is far below optimal. However, users can engage in further accelerator microarchitecture explorations based on ARAPrototyper.

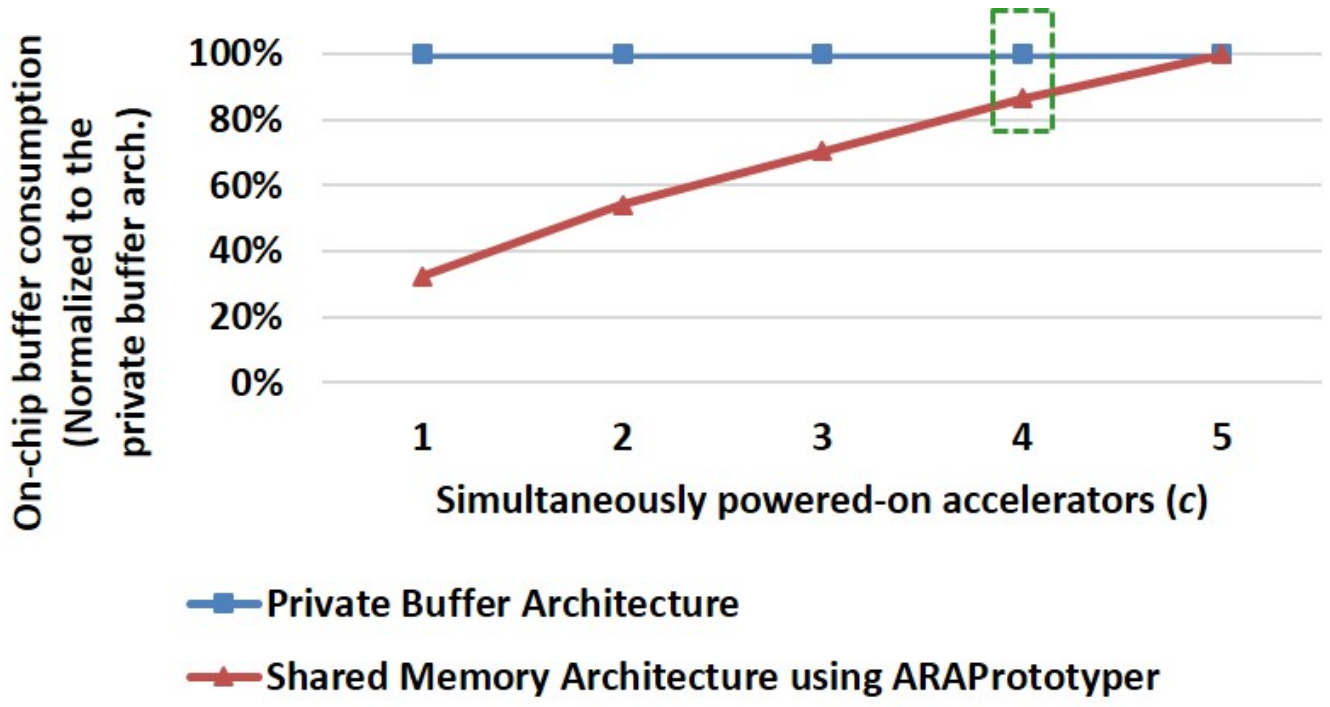

*Fig. 12:* Buffer consumption: private vs. shared buffer arch.

### *E. Design Space Exploration*

To demonstrate the capability and usage of ARAPrototyper, we conduct the following case studies for ARA design explorations.

*1) Private vs. Shared Buffer Architecture:* First, users can configure the ARAPrototyper to achieve either a private or shared buffer architecture (as explained in Section III-A1). To demonstrate this, we use an example ARA with a total of five accelerators. In the private buffer architecture, each accelerator needs its own buffer resources, regardless of how many of the accelerators are simultaneously powered on dynamically. While in the shared one, with the maximum number of simultaneously powered-on accelerators in mind, we can allocate the minimum buffer resources to support any

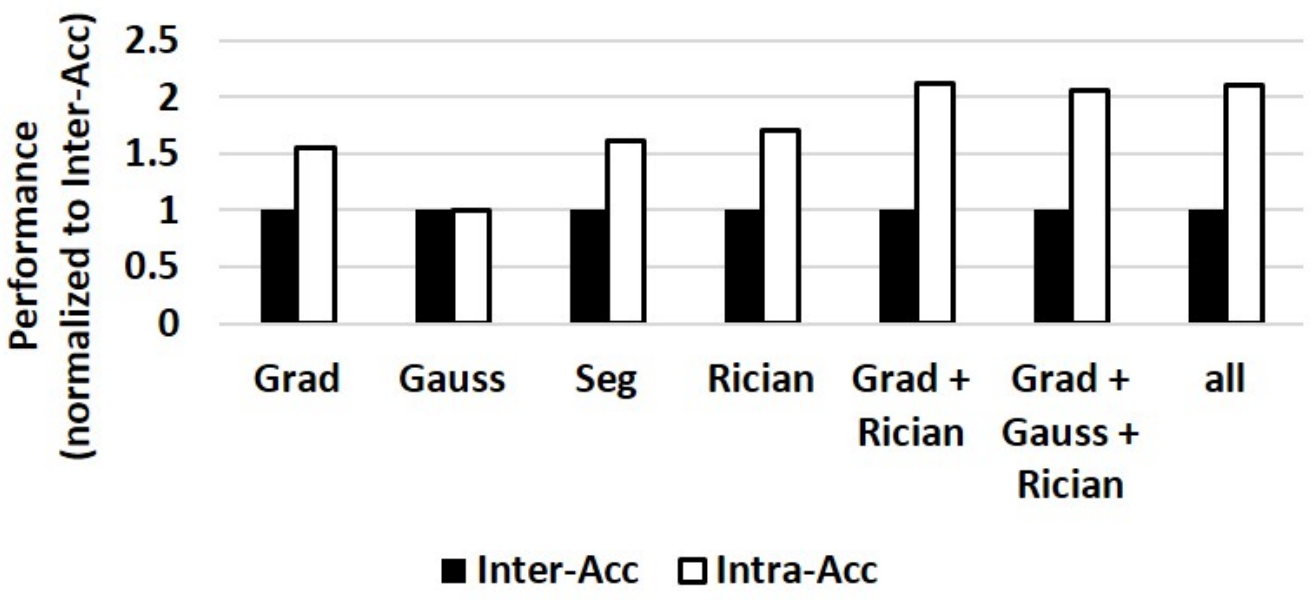

**(a)**

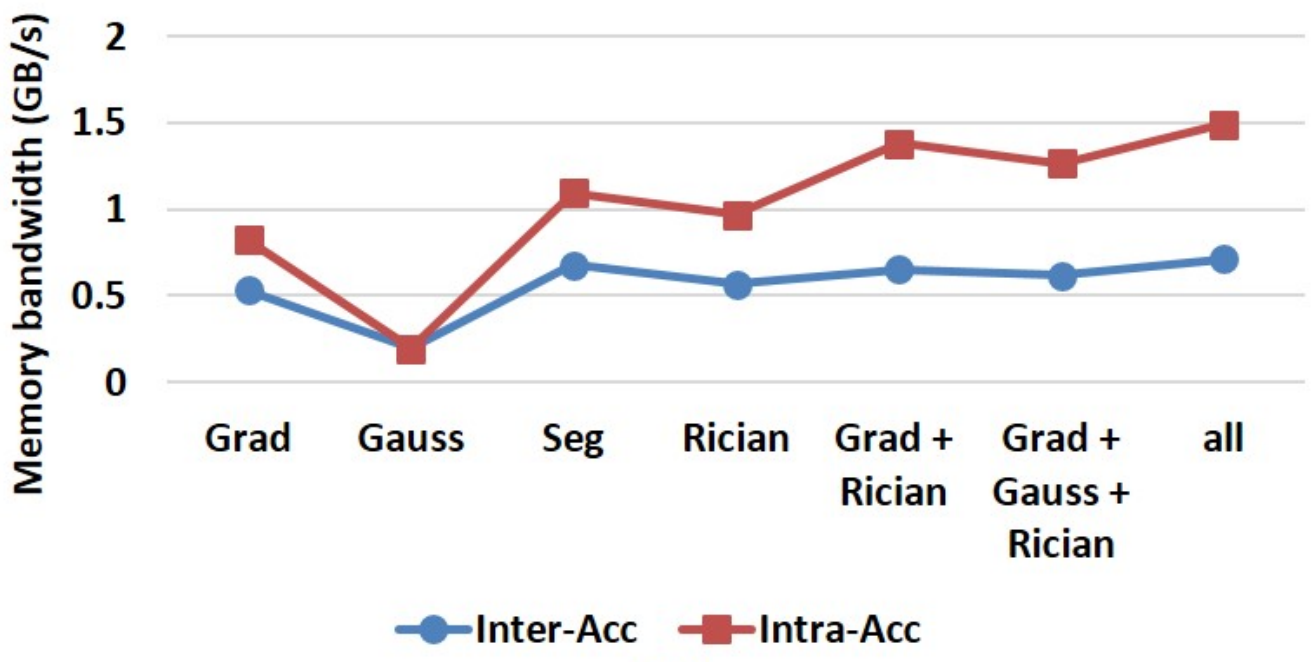

**(b)**

*Fig. 13:* Evaluation on (a) performance and (b) memory bandwidth between Inter-Acc and Intra-Acc interleaving networks.

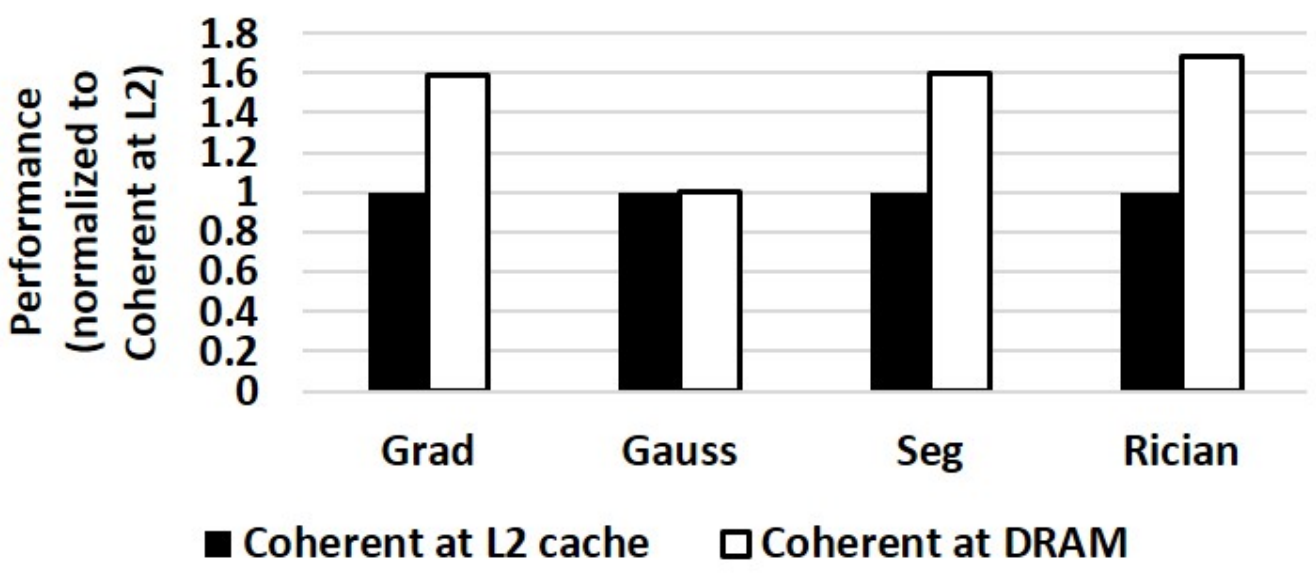

**(a)**

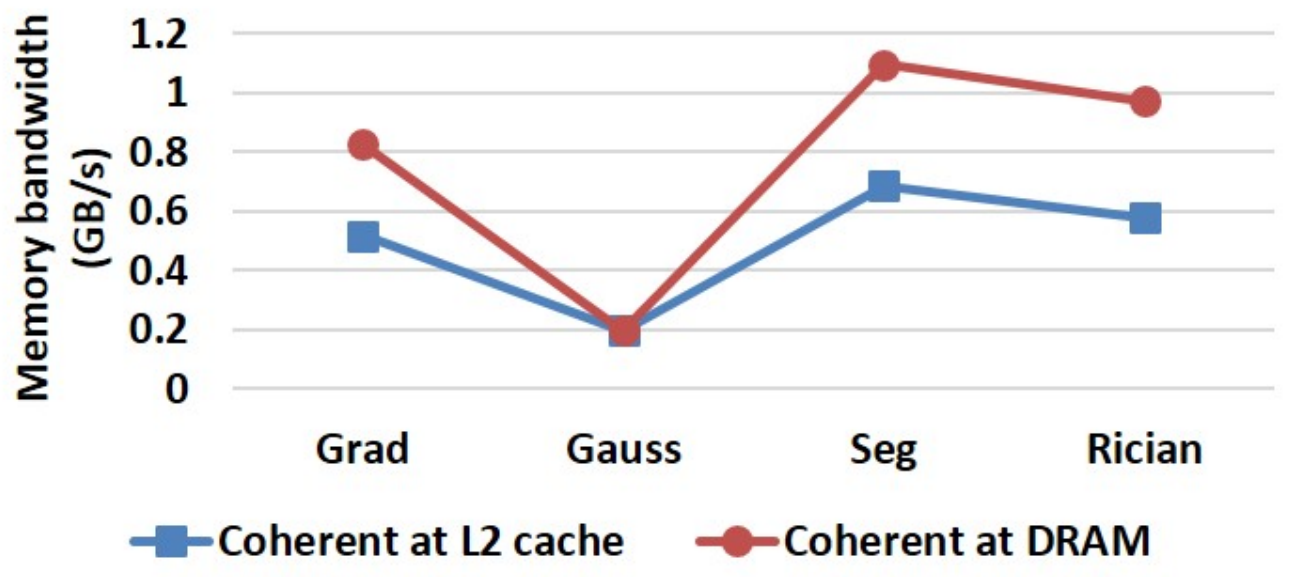

**(b)**

*Fig. 14:* Evaluation on (a) performance and (b) memory bandwidth for different coherency choices.

combination of accelerators running. Figure 12 shows that the shared buffer architecture can use much less physical buffer resources (thus less area and power) when not all accelerators are running simultaneously. On the other hand, if the shared buffer architecture is designed to support at most four simultaneous accelerators, but users need to run five tasks, then it would degrade the performance by 12.6% compared to the private buffer architecture (with 15.6% less buffer resources) based on our profiling.

*2) Interleaved Network: Inter-Acc vs. Intra-Acc:* Second, ARAPrototyper provides the flexibility for users to evaluate different interconnects between buffers and DRAM. Users can (statically) configure this interconnect to interleave inter-accelerators to achieve fairness among accelerators or within an accelerator to achieve better performance for the accelerator, as explained in Section III-A2. Figure 13(a) presents the performance of inter-accelerator and intra-accelerator interleaved networks for our medical imaging accelerators. As discussed in Section III-A2, the intra-accelerator interleaving can prevent the case in which all long-burst requests from the accelerator are issued to the same DMACs.

To gain more insights into this performance speedup, we further compare the achieved bandwidth of both cases; this can be obtained using the added performance counters in ARAPrototyper. As shown in Figure 13(b), intra-accelerator interleaving can achieve better bandwidth than inter-accelerator interleaving in our case, and thus achieves better performance. We can also observe that the available memory bandwidth is not the performance bottleneck. When we launch two, three, or four accelerators simultaneously, the utilized memory bandwidth still increases. Note that $gaussian$ is a special case since it only fetches four pages of data, and thus the impact is negligible.

*3) Coherency Choices:* Third, ARAPrototyper provides the flexibility for coherency choices at either LLC or DRAM depending on the application locality. Figure 14(a) presents the performance of both coherency choices. In our case, coherency at DRAM achieves up to 1.7X performance speedup compared to coherency at LLC. The major reason is that our medical imaging applications behave in a streaming fashion and have poor locality at LLC. Another reason is due to current Zynq board limitation, where the LLC has only one port while DRAM has four ports. As a result, coherency at DRAM can achieve higher bandwidth, as shown in Figure 14(b).

*4) Impact of TLB Sizes:* Fourth, The ARAPrototyper provides the flexibility for users to configure different TLB sizes. In addition, we also provide performance counters in ARAPrototyper to get the number of TLB accesses and misses for further performance analysis. A TLB miss handling (in software) penalty can also be collected in our system software stack. Figure 15(a) and Figure 15(b) present the TLB miss rate and TLB miss handling penalty (in terms of percentage of total execution time) for different TLB sizes. In our case, we choose 32K TLB entries in the design since the TLB miss rate and miss handling penalty will stop to shrink after this point. Another point we want to mention is that TLB misses can cause up to a penalty of 24% of the whole execution time in an ARA due to the streaming access behavior and accelerated computation. Therefore, it needs more attention to address translation support when designing an ARA compared to a general-purpose CPU.

*5) Accelerator Microarchitecture Exploration:* Finally, without loss of generality, users can conduct accelerator microarchitecture explorations such as 1) algorithm-level changes, and 2) HLS pragmas tuning. Actually, ARAPrototyper makes the accelerator microarchitecture explorations easier by providing more profiling statistics through performance counters and pointing out the optimization directions.

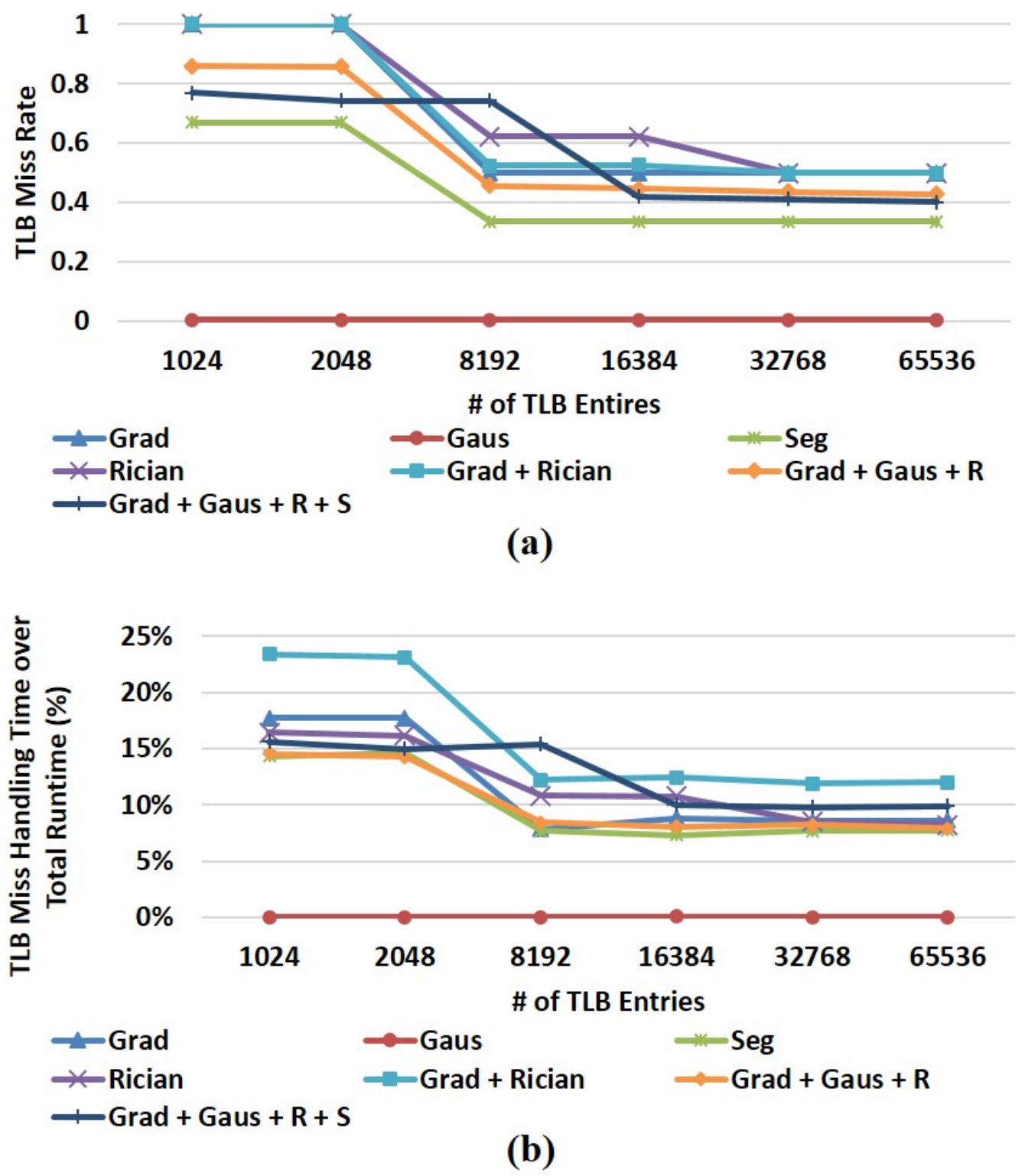

*Fig. 15:* The impact of TLB sizes on: (a) TLB miss rates and (b) TLB miss penalty over total runtime.

In this subsection we demonstrate a data reuse optimization for accelerators motivated by the profiled low computation percentage of total execution time in initial accelerator design. As shown in Figure 16(a), in our initial accelerator design, before data reuse optimization, the computation ratio is below 40% for most accelerators, which suggests the accelerators are not fully utilized but are waiting for data. Therefore, we apply the data reuse optimization presented in [43] to the accelerators. After this optimization, the computation ratio can be significantly increased, in most of the cases above 80%. As shown in Figure 16(b), the data reuse optimization can achieve up to 6X performance speedup.

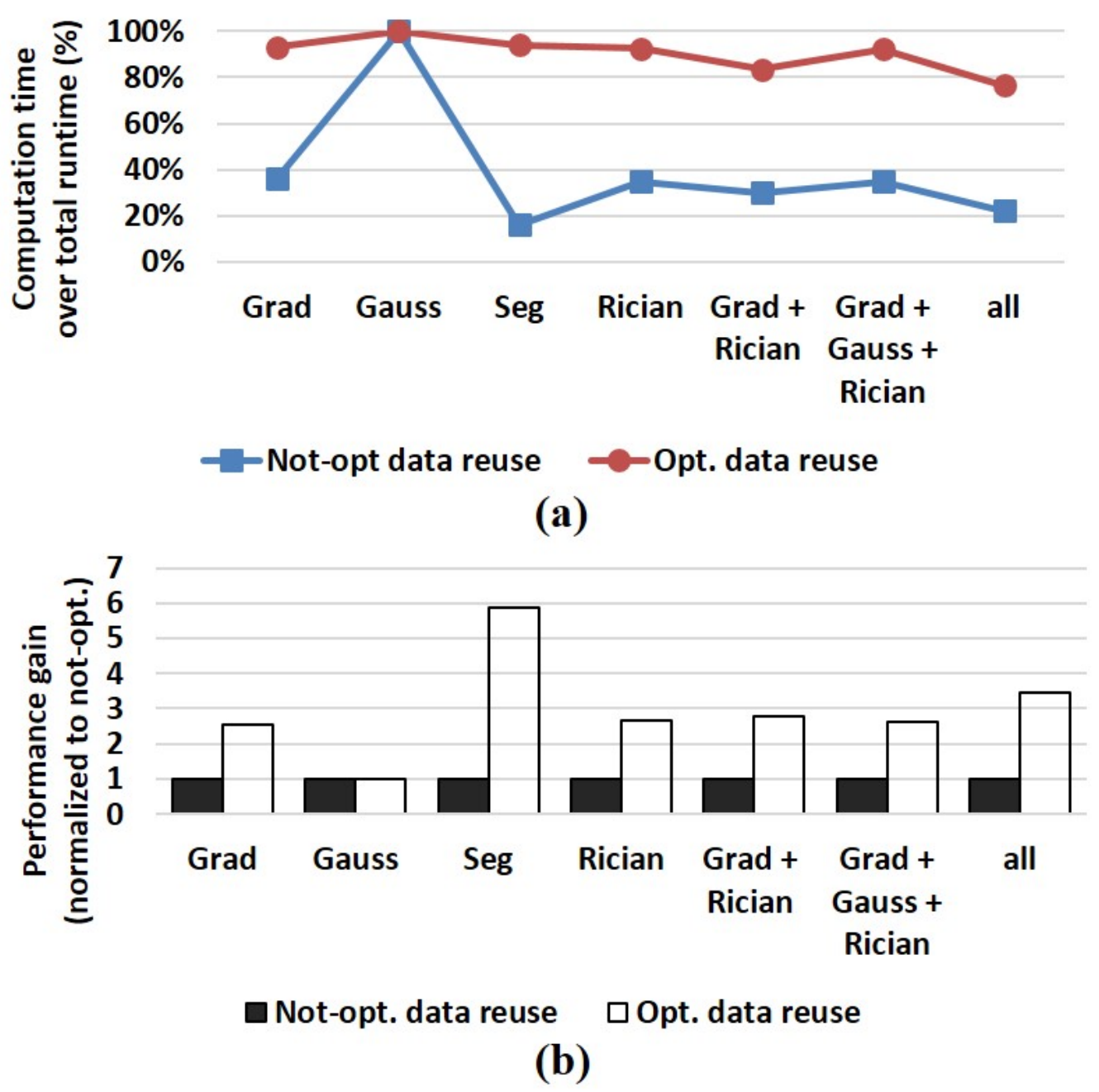

*Fig. 16:* Evaluation of accelerator data reuse optimization [43]: (a) the ratio of computation in total runtime; (b) performance speedup.

## VII. Conclusions

In this work we designed and implemented ARAPrototyper to enable rapid design space explorations for ARAs in FPGA prototypes with manageable efforts. Designers can easily integrate their HLS-compatible accelerator designs into our reusable baseline prototype for a few lines of code, and customize their own ARAs with up to 33 lines of XML code. The memory system of our ARA prototype is highly customizable and enables numerous design space explorations with insights provided by our added performance counters. Furthermore, we provide user-friendly APIs and the underlying system software stack for users to quickly develop their applications and deploy them seamlessly on our prototype. Finally ARAPrototyper achieves 4,000X to 10,000X faster evaluation time than full-system simulations. We believe that ARAPrototyper can be an attractive alternative for ARA design and evaluation.